\documentclass{amsart}
\usepackage[dvips]{color}
\usepackage{lmodern}
\usepackage[T1]{fontenc}
\usepackage{stmaryrd}
\usepackage{amsmath}
\usepackage[foot]{amsaddr}

\begin{document}

\title[7D Einstein and 6D Weyl gravitons]
{One-loop divergences in 7D Einstein and 6D Conformal Gravities}
\author{R. Aros $^{\ddag}$, F. Bugini $^{\S}$ and D.E. Diaz $^{\dag}$}
\address{${\ddag}$ Departamento de Ciencias F\'isicas, Universidad Andres Bello, Sazi\'e 2212, Piso 7, Santiago,
Chile}
\address{${\S}$ Departamento de Matem\'atica y F\'isica Aplicadas, Universidad Cat\'olica de la Sant\'isima Concepci\'on, Alonso de Ribera 2850, Concepci\'on, Chile}
\address{${\dag}$ Departamento de Ciencias F\'isicas, Universidad Andres Bello, Autopista Concepcion-Talcahuano 7100, Talcahuano, Chile}
\email{ \hspace*{0.0cm}raros@unab.cl , fbugini@ucsc.cl , danilodiaz@unab.cl }
\begin{abstract}
The aim of this note is to unveil a striking equivalence between the one-loop divergences in 7D Einstein and 6D Conformal Gravities.\\
The particular combination of 6D pointwise Weyl invariants of the 6D Conformal Gravity corresponds to that of Branson's Q-curvature and can be written solely in terms of the Ricci tensor and its covariant derivatives.
The quadratic metric fluctuations of this action, 6D Weyl graviton, are endowed with a sixth-order kinetic operator that happens to factorize on a 6D Einstein background into product of three shifted Lichnerowicz Laplacians. We exploit this feature to use standard heat kernel techniques and work out in one go the UV logarithmic divergences of the theory that contains in this case the four Weyl anomaly coefficients.\\
In a seemingly unrelated computation, we determine the one-loop IR logarithmic divergences of 7D Einstein Gravity in a particular 7D Poincar\'e-Einstein background that is asymptotically hyperbolic and has the above 6D Einstein manifold at its conformal infinity or boundary.\\
We show the full equivalence of both computations, as an outgrowth of the IR/UV connection in AdS/CFT correspondence, and in this way the time-honoured one-loop calculations in Einstein and higher-derivative gravities take an interesting new turn.
\end{abstract}


\maketitle

\section{Introduction}
\qquad\\
Higher-curvature gravities naturally turned up in response to the nonrenormalizable UV divergences of Einstein gravity~\cite{tHooft:1974toh,Deser:1974cz,Goroff:1985sz}. In four dimensions, attention was paid to quadratic gravities
once their perturbative renormalizability was established~\cite{Stelle:1976gc}. Nowadays, quadratic and higher-curvature gravities can be embedded into a fundamental theory like string- or M-theory and the notorious lack of unitarity can be attributed to an artifact of the truncation of the otherwise ghost-free  UV completion (see e.g.~\cite{Alvarez-Gaume:2015rwa}).

Prompted as well by developments in string theory, the study of higher-derivative gravities in dimensions larger that four also gained a renewed interest.
Take, for example, six dimensions where things are somewhat different. There is only one Gauss-Bonnet term that cannot absorb the dependence on the Riemann or Weyl tensor of the one-loop divergences, so that pure Einstein gravity turns out to be nonrenormalizable already at one-loop~\cite{PvN77,Deser:2016tgn}. The analogous role of 4D quadratic gravities is played now by six-derivative gravities containing cubic powers of Ricci and Riemann tensors. In particular, 6D Weyl or Conformal Gravities built up out of the three 6D pointwise Weyl invariants are indeed one-loop renormalizable by power-counting arguments.

The focus of our present work is a particular 6D Conformal Gravity, the study of the precise structure of its one-loop UV-log divergences -initiated in~\cite{Pang:2012rd} and complemented in~\cite{Beccaria:2017lcz}- and, eventually, the way they fit into the AdS/CFT correspondence. There are at least two features, as explained in~\cite{Beccaria:2017lcz}, that single out this 6D Conformal Gravity:
(i) it vanishes on a Ricci flat background,
and
(ii) it admits a (2,0) supersymmetric extension.
The first one is best known to conformal geometers for this is a crucial property of Branson's Q-curvature, the quantity that arises within AdS/CFT as the volume anomaly of (asymptotically AdS) Poincar\'e-Einstein metrics~\cite{GZ03}. The second property is related to the fact that the very same combination of pointwise Weyl invariants appears in the accumulated $b_6$ heat coefficient for the free (2,0) tensor multiplet~\cite{Bastianelli:2000hi}.
\\
The precise combination of pointwise Weyl invariants that makes up the 6D Conformal Gravity under consideration is the following
\begin{equation}
S_{CG}=\int d^6x \sqrt{g} \left[ Ric\nabla^2Ric -\frac{3}{10}R\nabla^2R+2 Riem Ric^2-R Ric^2 + \frac{3}{25}R^3\right]
\end{equation}
The one-loop partition function for the corresponding 6D Weyl graviton can be obtained by integrating out the quadratic metric fluctuations after fixing the Feynman-de Donder gauge and taking into account the ghost contribution. The major technical difficulty in doing so is posed by the six-order kinetic operator acting on the transverse-traceless metric fluctuations. However, restricting to an Einstein background, the computations are greatly simplified, and we end up with the following quotient of functional determinants of (minimal) second-order differential operators, simple-shifted Lichnerowicz Laplacians,
\begin{equation}
Z^{^{1-loop}}_{_{Weyl}}=\left[\frac{\det\left\{{\Delta}_L^{(1,\bot)}-\frac{1}{3}R\right\}\cdot\det\left
\{{\Delta}_L^{(0)}-\frac{1}{5}R\right\}}{\det\left\{{\Delta}_L^{(2,\bot\top)}-\frac{1}{3}R\right\}\cdot
\det\left\{{\Delta}_L^{(2,\bot\top)}-\frac{1}{5}R\right\}\cdot\det\left\{{\Delta}_L^{(2,\bot\top)}
-\frac{2}{15}R\right\}}\right]^{1/2}
\end{equation}\\
This factorized form first appeared in the physics literature in~\cite{Pang:2012rd}, but it was incorrectly claimed to hold up only on symmetric Einstein manifolds. An extension of the factorization to Ricci flat manifolds was exploited later on by~\cite{Beccaria:2017lcz}. In greater generality, the factorization of the second metric variation of the (critical) Q-curvature on a generic Einstein manifold was established by~\cite{Mat14} using the Fefferman-Graham ambient metric construction. For the sake of completeness, we give a detailed proof of the factorization for the current 6D case in appendix~\ref{app.A}.

The structure of the UV-log divergences of any 6D Weyl invariant action is dictated by the trace (or Weyl or conformal) anomaly (see e.g.~\cite{Bastianelli:2000hi}):
\begin{equation}
{\mathcal A}\,=\,-{a}\,E_6\,+\,{c_1}\,I_1\,+\,{c_2}\,I_2\,+\,{c_3}\,I_3
\end{equation}
The restriction to symmetric Einstein spaces, such as $S^6,\,S^2\times S^4,\,S^3\times S^3,\,S^2\times S^2\times S^2$ where the Weyl tensor is covariantly constant,  grants access to the coefficient of the Pfaffian $E_6$ but forces a linear relation between the pointwise  Weyl invariants $5I_3=32I_1-8I_2$. As a consequence, there is not enough information to disentangle the four anomaly coefficients from only three independent terms
\begin{equation}
{\mathcal A}\,=\,-{a}\,E_6\,+\,{(c_1+4c_2)}\,I_1\,+\,{(c_3-\frac{5}{8}c_2)}\,I_3
\end{equation}
This restricted approach to the determination of the UV-log divergences of the one-loop effective action for the 6D Weyl graviton was carried out in~\cite{Pang:2012rd} by explicit computation of the eigenvalues and degeneracies of the second-order differential operators that enter the functional determinants. The partial information obtained by this procedure was then the following
\begin{equation}
a\,=\,\frac{601}{2016}\;,\quad c_{1}+4c_{2}=\frac{5633}{105}\;,
\quad c_{3}-\frac{5}{8}c_{2}=-\frac{35543}{5040}~.
\end{equation}\\
More recently, the authors of~\cite{Beccaria:2017lcz} considered restriction to a Ricci-flat, but not conformally flat, background. Ricci flatness forces two linear relations between the four terms of the anomaly basis, namely, $E_6=64I_1+32I_2$ and $I_3=4I_1-I_2$, so that there are only two independent terms in the anomaly, say
\begin{equation}
{\mathcal A}\,=\,-{[a- \frac{1}{192}(c_1+4c_2)]}\,E_6\,+\,{[c_3+ \frac{1}{6}(c_1-2c_2)]}\,I_3
\end{equation}
The coefficients of these two independent combinations were obtained by explicit evaluation of the accumulated $b_6$ heat kernel coefficients of all the second-order kinetic operators involved. When combined with the previous partial results from symmetric Einstein manifolds, the first term brings in no new information but the second one allows the complete determination of $c_3$ and, in consequence, of $c_1$ and $c_2$.  As a result, both computations nicely complement each other to produce
the full set of central charges

\begin{equation}
a\,=\,\frac{601}{2016}\;,\quad c_{1}=\frac{1507}{45}\;,
\quad c_{2}=\frac{635}{126}\;,\quad c_{3}=-\frac{1639}{420}~.
\end{equation}\\

Let us now turn to the central question of interest in this paper, namely, deciphering a hologram: somewhat unexpectedly, this 6D Conformal Gravity computation has a 7D bulk counterpart within AdS/CFT correspondence. There is a kinematic relation between the one-loop partition functions of the bulk Einstein graviton and the boundary Weyl graviton~\cite{Giombi:2013yva}

\begin{equation}\label{holo}
\Large{\frac{Z^{^{1-loop,-}}_{_{Einstein}}}{Z^{^{1-loop,+}}_{_{Einstein}}}\,=\,Z^{^{1-loop}}_{_{Weyl}}}
\end{equation}\\
The bulk side contemplates the ratio of the functional determinants of the kinetic operator of the bulk field computed with standard and alternate boundary conditions, whereas the boundary side involves the functional determinant of the kinetic operator of the induced field. This kind of {\it holographic formula}  was obtained via a rather circuitous route within AdS/CFT correspondence, in connection with a class of RG flows triggered by double-trace deformations of the CFT~\cite{Gubser:2002zh,Gubser:2002vv,Hartman:2006dy}.
Full match for a massive scalar in Euclidean AdS bulk was first shown in~\cite{Diaz:2007an}, and by now there are plenty of extensions: an incomplete list includes fields with nonzero spin (Dirac, MHS, etc.) and quotients of AdS space (thermal AdS, BTZ, singular AdS, etc.)~\cite{Diaz:2008hy}-\cite{Bugini:2018def}.\\
Let us stress that although the 1-loop holographic relation between fluctuation determinants is valid for generic spacetime dimensions, there is an essential difference at tree level. The boundary Weyl graviton action, induced by the bulk Einstein graviton, around flat space takes the following form in terms of the flat-space Laplacian $\partial^2=\Box$ and the linearized Weyl tensor $w$
 \begin{equation}
S_{_{Weyl}}=\int d^dx\,w\,\Box^{\frac{d}{2}-2} w~.
\end{equation}
This clearly corresponds to the linearization of the full non-linear Lagrangian given by the Weyl tensor squared $W^2$ in 4D and of the $I_3\sim W\Box W$ Weyl invariant in 6D. By contrast, in odd dimensions the induced action corresponds to a non-local kernel given by a half-integer power of the Laplacian. In particular, the induced Weyl or conformal invariant action in 3D is not of local Chern-Simons type but has a parity-even non-local action and no effective degrees of freedom (see, e.g.,~\cite{Beccaria:2016tqy}). There are no log-divergences or Weyl anomaly on closed manifolds in this case, but still the holographic formula can be studied on thermal quotients of AdS and in connection with holographic entanglement entropy. In the present work, we restrict to even dimensions and focus on the matching of log-divergences. 

The precise correspondence between one-loop partition functions for the 5D Einstein graviton and 4D Weyl graviton has already been established: although the one-loop divergences of 4D Weyl gravity have long been known~\cite{Fradkin:1981hx,Fradkin:1981iu},  the bulk counterpart has recently been given a successful treatment~\cite{Acevedo:2017vkk}. In turn, the 7D holographic counterpart, to date, only accounts for the type-A anomaly coefficient by explicit evaluation of the corresponding functional determinants on hyperbolic space~\cite{Giombi:2013yva}.

Our present contribution is twofold. First, we obtain, in one go, all anomaly coefficients by extending the boundary computation to a generic Einstein background, exploiting the factorization into shifted Lichnerowicz operators~\cite{Mat14} and making use of the corresponding explicit form of $b_6$ heat coefficient for the symmetric transverse-traceless two-tensor field~\cite{Liu:2017ruz}. Second, we extend the bulk computation to an asymptotically AdS Poincar\'e-Einstein metric, with the above Einstein manifolds on its conformal boundary, in order to compute the log-IR divergence of the 7D Einstein graviton. We do verify the complete agreement with the previous boundary computation of the log-UV divergence of the 6D Weyl graviton, in accordance with expectations from the AdS/CFT dictionary at one loop and as an outgrowth of the IR-UV connection~\cite{Susskind:1998dq}.

\qquad\\
\section{6D Conformal Gravity: UV divergence}
\qquad\\

As already stated, the one-loop partition function of the 6D Conformal Gravity of interest is given by the following quotient of functional determinants of
fluctuation operators when restricted to an Einstein background

\begin{equation}
Z^{^{1-loop}}_{_{Weyl}}=\left[\frac{\det\left\{{\Delta}_L^{(1,\bot)}-\frac{1}{3}R\right\}\cdot\det\left
\{{\Delta}_L^{(0)}-\frac{1}{5}R\right\}}{\det\left\{{\Delta}_L^{(2,\bot\top)}-\frac{1}{3}R\right\}\cdot
\det\left\{{\Delta}_L^{(2,\bot\top)}-\frac{1}{5}R\right\}\cdot\det\left\{{\Delta}_L^{(2,\bot\top)}
-\frac{2}{15}R\right\}}\right]^{1/2}
\end{equation}\\
\\
In six dimensions, the UV logarithmic divergence is given by the ``accumulated'' $b_6$ heat coefficient. The scalar and vector inputs have been known for quite a while, whereas the coefficient for the rank 2 symmetric traceless unconstrained tensor has only recently been reported~\cite{Liu:2017ruz}. We write down the latter in the full A-basis of curvature invariants\footnote{We refer to~\cite{Bastianelli:2000hi} for notational conventions.} for future reference

\begin{align}
7!\cdot\overline{{b}}_6\{{\Delta}_L^{(2,\top)}\}\,=&\, -312A_1 - 584A_2 - 4552A_3 + 368A_4 + 544A_5 - 1064A_6
\\
\nonumber
\\
\nonumber & + 9920A_7 - 416A_8 + 1416A_9 - \frac{560}{9}A_{10} + \frac{7952}{3}A_{11} + \frac{2968}{3}A_{12}\\
\nonumber
\\
\nonumber &\, - \frac{117056}{9} A_{13} - \frac{16528}{3}A_{14} - \frac{29216}{3}A_{15} - \frac{1388}{9}A_{16} + \frac{49984}{9}A_{17}
\end{align}\\
\\
We can now apply our shortcut route by restricting to an Einstein metric without spoiling the independence of the four Weyl-invariant terms in the 6D Weyl anomaly basis~\footnote{This shortcut route to the Weyl anomaly has already been noticed in~\cite{Bugini:2016nvn} and successfully put into use by~\cite{Beccaria:2017dmw,Acevedo:2017vkk,Bugini:2018def}.}. It is enough then to keep track on the numerical coefficients in front of $\{A_5, A_9, A_{10}, A_{11},A_{12},A_{13},A_{14},A_{15},A_{16},A_{17}\}$\footnote{$A_9$ is equivalent to $-A_5$, up to a trivial total derivative.}. In addition, due to the constant shifts in the Lichnerowicz Laplacians, we need convolution with exponentials of the constant Ricci scalar. Finally, we need to subtract longitudinal and trace components of the tensor and vector operators, which turns out to be an easy task when written in terms of Lichnerowicz Laplacians (see e.g.~\cite{Percacci}, eqns. 5.117 and 5.123). \\
\\
In all, we obtain the following\footnote{We notice a misprint in the coefficient of $A_{13}$ for the scalar in~\cite{Liu:2017ruz}, but fortunately it does not alter the subsequent results therein.}
\begin{align}
7!\cdot\overline{{b}}_6\{{\Delta}_L^{(0)}-R/5\}\,=\,&\, 9A_5 + 12A_9 + \frac{9107}{225}A_{10} - \frac{14}{3}A_{11} + \frac{154}{15}A_{12}\\
\nonumber
\\
\nonumber &\, - \frac{208}{9}A_{13} + \frac{64}{3}A_{14} - \frac{16}{3}A_{15} + \frac{44}{9}A_{16} + \frac{80}{9}A_{17}
\end{align}\\
\begin{align}
7!\cdot\overline{{b}}_6\{{\Delta}_L^{(1,\bot)}-R/3\}\,=&\, -67A_5 - 108A_9 + \frac{245}{9}A_{10} +
\frac{1274}{3}A_{11} - 168A_{12}\\
\nonumber
\\
\nonumber &\, - \frac{7592}{9}A_{13} - \frac{16}{3}A_{14} + \frac{1348}{3}A_{15} - \frac{536}{9}A_{16} - \frac{1112}{9}A_{17}
\end{align}\\
\begin{align}
7!\cdot\overline{{b}}_6\{{\Delta}_L^{(2,\bot\top)}-R/3\}\,=&\, 602A_5 + 1512A_9 - \frac{1456}{9}A_{10} + \frac{6692}{3}A_{11} + 2730A_{12}\\
\nonumber
\\
\nonumber &\, - \frac{109256}{9}A_{13} - \frac{16576}{3}A_{14} - \frac{30548}{3}A_{15} - \frac{896}{9}A_{16} + \frac{51016}{9}A_{17}
\end{align}\\
\begin{align}
7!\cdot\overline{{b}}_6\{{\Delta}_L^{(2,\bot\top)}-R/5\}\,=&\, 602A_5 + 1512A_9 + \frac{9688}{225}A_{10} + \frac{6692}{3}A_{11} + \frac{30926}{15}A_{12}\\
\nonumber
\\
\nonumber &\, - \frac{109256}{9}A_{13} - \frac{16576}{3}A_{14} - \frac{30548}{3}A_{15} - \frac{896}{9}A_{16} + \frac{51016}{9}A_{17}
\end{align}\\
\begin{align}
7!\cdot\overline{{b}}_6\{{\Delta}_L^{(2,\bot\top)}-2R/15\}\,=&\, 602A_5 + 1512A_9 + \frac{20972}{225}A_{10} + \frac{6692}{3}A_{11} + \frac{8638}{5}A_{12}\\
\nonumber
\\
\nonumber &\, - \frac{109256}{9}A_{13} - \frac{16576}{3}A_{14} - \frac{30548}{3}A_{15} - \frac{896}{9}A_{16} + \frac{51016}{9}A_{17}
\end{align}\\
The ``accumulated'' heat coefficient in the reduced A-basis then reads\\
\begin{align}
&\, 1864A_5 + 4632A_9 - \frac{20972}{225}A_{10} + 6272A_{11} + \frac{100156}{15}A_{12} \\
\nonumber
\\
\nonumber &\, - 35552A_{13} - 16576A_{14} - 244A_{16} + 17120A_{17}
\end{align}\\
To read off the 6D Weyl anomaly ${\mathcal A}$, that is, the coefficient of the UV-log divergence, it is convenient to go to the basis where one trades the Euler density by the Q-curvature, which reduces to a multiple of the Ricci scalar cubed in the present case of an Einstein metric
\begin{eqnarray}
{\mathcal A}&=&-{a}\,E_6\,+\,{c_1}\,I_1\,+\,{c_2}\,I_2\,+\,{c_3}\,I_3 \\
\nonumber\\\nonumber
\qquad\quad&=&-48\,a\,{\mathcal Q}_6+(c_1-96a)I_1+(c_2-24a)I_2+(c_3+8a)I_3\\
\nonumber\\\nonumber
\qquad\quad&=&-16\,a\,R^3/75+(c_1-96a)I_1+(c_2-24a)I_2+(c_3+8a)I_3
\end{eqnarray}

\[
\begin{array}{|r c| c| c| c| c| c|} \hline
& \mbox{Curvature invariant } & {\mathcal Q}_6=R^3/225 & I_1 & I_2 &  I_3 \\
\hline {A}_{10}\quad\vline & {R}^{\,3}   &225 &  -&- &- \\
\hline {A}_{11}\quad\vline & {R}{R}ic^{\,2} & 75/2  & -&- &- \\
\hline {A}_{12}\quad\vline & {R}{R}iem^{\,2} &15  &20  &-5 &-5 \\
\hline {A}_{13}\quad\vline & {R}ic^{\,3} & 25/4& -& -& -\\
\hline {A}_{14}\quad\vline & {R}iem \, {R}ic^{\,2} &25/4 & -&- &-  \\
\hline {A}_{15}\quad\vline & {R}ic \, {R}iem^{\,2} & 5/2 &10/3  &-5/6 &-5/6 \\
\hline {A}_{16}\quad\vline & {R}iem^{\,3} & 1 & 4 & 0& -1 \\
\hline {A}_{17}\quad\vline & -{R}iem'^{\,3} &1 & -2& 1/4&1/4  \\
\hline {A}_{5}\quad\vline & |{\nabla}{R}iem|^{2}   &- &-32/3 &8/3 &5/3  \\
\hline
\end{array}
\]
\\
The dictionary in the table above allows us to collect all independent contributions

\begin{equation}
7!\cdot{\mathcal A} \,=\,-72120\,{\mathcal Q}_6 + 24544\,I_1 - 10660\,I_2 - 7648\,I_3
\end{equation}
\\
We finally find out the central charges in full agreement with~\cite{Pang:2012rd,Beccaria:2017lcz}

\begin{equation}
a\,=\,\frac{601}{2016}\;,\quad c_{1}=\frac{1507}{45}\;,
\quad c_{2}=\frac{635}{126}\;,\quad c_{3}=-\frac{1639}{420}~.
\end{equation}
\qquad\\

\section{7D Einstein Gravity: IR divergence}
\qquad\\
As anticipated by the AdS/CFT dictionary~\cite{Giombi:2013yva}, we now consider the one-loop partition function of the 7D Einstein graviton, that is, the functional determinant of the quadratic metric fluctuations of the Einstein-Hilbert action

\begin{equation}
S_{_{EH}}=-\frac{1}{2\kappa^2}\int d^7x \sqrt{\hat{g}_{PE}} \left[\hat{R}-2\hat{\Lambda}\right]
\end{equation}
This is a much more familiar computation that proceeds along standard lines, see e.g.~\cite{Percacci} and references therein, with the only unusual feature possibly being the dimensionality of spacetime we are considering, namely,  seven. The Feynman-de Donder gauge choice cancels out nonminimal terms in the Hessian and in the gauge-fixing term, which leads to significant simplification of the kinetic operators involved in the one-loop computation. The restriction to an Einstein bulk background further reduces the calculation to the following quotient of functional determinants of second-order operators, shifted Lichnerowicz Laplacians,

\begin{equation}
Z^{^{1-loop}}_{_{Einstein}}=\left[\frac{\det\left\{\hat{\Delta}_L^{(1,\bot)}-\frac{2}{7}\hat{R}\right\}}
{\det\left\{\hat{\Delta}_L^{(2,\bot\top)}-\frac{2}{7}\hat{R}\right\}}\right]^{1/2}
\end{equation}
\qquad\\
We need to study the IR-log divergence of this partition function stemming from the infinite volume of the bulk background. Heat kernel techniques will prove a convenient computational tool to account for the standard (Dirichlet) boundary conditions in $Z^{^{1-loop,+}}_{_{Einstein}}$, whereas the alternate (Neumann) boundary conditions in $ Z^{^{1-loop,-}}_{_{Einstein}}$ are reached by analytic continuation in the spectral parameter, that is, in the scaling dimension of the induced boundary fields~\footnote{For the type-A anomaly coefficient $a$, this has previously been exploited in~\cite{Giombi:2013yva}. We will confirm the validity of the analytic continuation for both type-A and type-B anomaly coefficients.}.

\subsection{Conformally flat case: hyperbolic space}
\qquad\\

The heat kernels for the transverse vector and for the transverse-traceless symmetric rank-two tensor in hyperbolic space has long been known~\cite{Camporesi:1993mz,Camporesi:1994ga,Gopakumar:2011qs}.
A remarkable feature, referred to as WKB exactness, is that after factorization of the exponential of a multiple of the Ricci scalar~\footnote{We set the radius of the hyperbolic space to 1, so that the bulk Ricci scalar is simply $\hat{R}=-42$.}, the trace only has finite many terms so that the combined contribution of tensor and vector is given by the following proper-time integral (in the standard route to the `log-dets')

\begin{align}
&\int_{0}^{\infty}\frac{dt}{t}\left\{\mbox{tr}\,e^{-(\hat{\Delta}_L^{(2,\bot\top)}+12)\,t}\,-\,\mbox{tr}
\,e^{-(\hat{\Delta}_L^{(1,\bot)}+12)\,t}\,\right\}
\\\nonumber
\\\nonumber
\sim&\int_{0}^{\infty}\frac{dt}{t^{9/2}}\left\{e^{-9t}\left[20 + 136\,t + \frac{256}{3}\,t^2\right] - e^{-16t}\left[ 6 + 24\,t + \frac{72}{5}\,t^2\right]\right\}
\end{align}\\
After evaluation of the proper-time integral in terms of gamma functions, we obtain the numerical factor corresponding to the one-loop effective Lagrangian on hyperbolic space that accompanies the volume, following the prescription of~\cite{ISTY99}. The volume anomaly is given by the Q-curvature so that we can directly read off the type-A Weyl anomaly coefficient, it coincides with the result obtained from the boundary computation on the round six-sphere

\begin{equation}
a\,=\,\frac{3005}{2\times 7!}=\frac{601}{2016}
\end{equation}\\
This holographic result in the conformally flat situation was first derived in~\cite{Giombi:2013yva} and, in fact,  established for the whole family of bulk gauge fields of higher spins dual to boundary conformal higher spins.

\subsection{Non-conformally flat case: Poincar\'e-Einstein metric}
\qquad\\

Now we need to deviate from conformal flatness by considering a more general Poincar\'e-Einstein metric, and we choose the particular one that has an Einstein metric on the conformal class of boundary metrics~\cite{Besse,FG12}. This bulk metric has a finite Fefferman-Graham expansion and is given by~\footnote{Here $\lambda\,=\,R/4d(d-1)$ is a multiple of the (necessarily constant) Ricci scalar $R$ of the (d-dimensional) boundary Einstein metric.}

\begin{equation}\label{PEE}
\hat{g}_{_{PE/E}}=\frac{dx^2+(1-{\lambda}\,x^2)^2{g_{_E}}}{x^2}
\end{equation}\\

 As the information on the conformally flat case has already been extracted, we only need to focus on the Weyl-tensor content of the heat coefficients when evaluated on this particular bulk background metric.

\[
\begin{array}{|r c| c| c| c| c|} \hline
& \mbox{Curvature invariant } & \hat{W}'^{\,3} & \hat{W}^{3} &  \hat{\Phi}_7 \\
\hline \widehat{A}_{12}\quad\vline & \widehat{R}\widehat{R}iem^{\,2}  & -42 & 21/2& -21/2\\
\hline \widehat{A}_{15}\quad\vline & \widehat{R}ic \, \widehat{R}iem^{\,2} & -6 & 3/2& -3/2\\
\hline \widehat{A}_{16}\quad\vline & \widehat{R}iem^{\,3} & -6 & 5/2&-3/2  \\
\hline \widehat{A}_{17}\quad\vline & -\widehat{R}iem'^{\,3} & 1/2 & -3/8 & 3/8  \\
\hline \widehat{A}_{5}\quad\vline & |\hat{\nabla}\widehat{R}iem|^{2} &8 &-2 &3  \\
\hline
\end{array}
\]
\\
The above table contains the only contributions from the general A-basis of curvature invariants. We choose
the 7D Fefferman-Graham invariant $\hat{\Phi}_7$ because it ``descends'' directly to the corresponding 6D $\Phi_6$
when reading off the holographic Weyl anomaly (or, equivalently, the logarithmic IR divergence) according to the simple prescription of~\cite{Bugini:2016nvn}. Evaluated on an Einstein metric, the Fefferman-Graham invariant~\cite{FG12} can be readily written in terms of the Weyl and the Schouten tensors $\Phi=|\nabla W|^2+16PW^2$, in the 6D Einstein boundary we then have $\Phi_6=|\nabla W|^2+\frac{4}{15}RW^2$ and in the 7D Poincar\'e-Einstein bulk (with unit AdS radius), $\hat{\Phi}_7=|\hat{\nabla} \hat{W}|^2-8\hat{W}^2$. Furthermore, in our particular Poincar\'e-Einstein metric (eqn.~\ref{PEE}), $\hat{\Phi}_7$ becomes proportional to the boundary $\Phi_6$ in exactly the same manner that the cubic bulk contractions $\hat{W}^{' 3}$ and $\hat{W}^3$ are related to their boundary counterparts. This is essentially the key observation of~\cite{Bugini:2016nvn}: being all on equal footing, their contributions to the 6D holographic Weyl anomaly are obtained by simply replacing the bulk invariants with the corresponding boundary ones.\\
The Weyl tensor comes into play starting with the $\hat{\bar{b}}_4$ heat coefficient. Upon convolution with the constant shift proportional to the Ricci scalar, it will also contribute to the $\hat{\bar{b}}_6$ coefficient. On top of that, we will need as well to factor out the exponentials in the Ricci scalar as demanded by the WKB exactness of the heat kernel.

For the transverse vector, the relevant heat coefficients can be worked out in 7D starting with the generic expressions reported in~\cite{Bastianelli:2000hi} for the unconstrained vector~\footnote{There is a misprint in the $b_6$ in arbitrary dimension $d$ reported in~\cite{Bastianelli:2000hi}, appendix~\ref{app.C}: the coefficient of the $A_{16}$ should read $(\frac{44d}{9}-84)$ instead of $(\frac{46d}{9}-84)$. In particular, at $d=6$ it must agree with the $-\frac{164}{3}$ of eqn.(2.21) therein.} and subtracting the scalar longitudinal component.
In terms of unconstrained vector and scalar heat kernels we have\\
\begin{equation}
\mbox{tr}\,e^{-(\hat{\Delta}_L^{(1,\bot)}\, + \,12)\,t}\,=\,e^{-16t}\left\{e^{4t}\,\mbox{tr}\,e^{-\hat{\Delta}_L^{(1)}\,t}
\,-\,e^{4t}\,\mbox{tr}\,e^{-\hat{\Delta}_L^{(0)}\,t}\right\}
\end{equation}
The Weyl content is contained, after convolution with the exponentials, in the following combination of heat coefficients for the Lichnerowicz Laplacians
\begin{align}
\mbox{tr}\,e^{-(\hat{\Delta}_L^{(1,\bot)}+12)\,t}\,\sim\;e^{-16t}
&\left\{\hat{\overline{b}}_4\{\hat{\Delta}_L^{(1)}\}\cdot t^2\,+
\,\left[4\,\hat{\overline{b}}_4\{\hat{\Delta}_L^{(1)}\}+\hat{\overline{b}}_6\{\hat{\Delta}_L^{(1)}\}\right]\cdot t^3 \right.
\\\nonumber
\\\nonumber
&\left.-\hat{\overline{b}}_4\{\hat{\Delta}_L^{(0)}\}\cdot t^2\,-
\,\left[4\,\hat{\overline{b}}_4\{\hat{\Delta}_L^{(0)}\}+\hat{\overline{b}}_6\{\hat{\Delta}_L^{(0)}\}\right]\cdot t^3\right\}
\end{align}
\\
Below we collect the relevant terms of the corresponding heat coefficients
\begin{equation}
\hat{\overline{{b}}}_4\{\hat{{\Delta}}_L^{(0)}\}\,\sim\,~\frac{1}{180}\,\hat{W}^2
\end{equation}
\begin{equation}
\hat{\overline{{b}}}_4\{\hat{{\Delta}}_L^{(1)}\}\,\sim\,~-\frac{2}{45}\,\hat{W}^2
\end{equation}\\
\begin{equation}
7!\cdot\hat{\overline{{b}}}_6\{\hat{{\Delta}}_L^{(0)}\}\,\sim\, -3\hat{A}_5 + \frac{14}{3}\hat{A}_{12} - \frac{16}{3}\hat{A}_{15} + \frac{44}{9}\hat{A}_{16} + \frac{80}{9}\hat{A}_{17}
\end{equation}\\
\begin{equation}
7!\cdot\hat{\overline{{b}}}_6\{\hat{{\Delta}}_L^{(1)}\}\,\sim\, 35\hat{A}_5 - \frac{196}{3}\hat{A}_{12} + \frac{1316}{3}\hat{A}_{15} - \frac{448}{9}\hat{A}_{16} - \frac{952}{9}\hat{A}_{17}
\end{equation}
It is convenient to keep the $\hat{W}^2$ accompanying $t^2$ and to translate the contribution to the $t^3$ coefficient into the basis of cubic Weyl invariants by means of the identity $\hat{W}^2\,=\, \hat{W}'^{\,3}\,-\,\frac{1}{4}\hat{W}^{3}\,+\,\frac{1}{4}\hat{\Phi}_7$, which is valid for the 7D Poincar\'e-Einstein metric with unit AdS radius~\footnote{This identity holds modulo a total derivative. More precisely, it  follows from the total derivative $\hat{C}_7$ (cf.~\cite{Bastianelli:2000hi}) when evaluated in the 7D PE/E metric. By contrast, in 5D the coefficient of the $\hat{W}^2$ term vanishes and one obtains an identity between the three Weyl invariants $\hat{W}'^{\,3}\,-\,\frac{1}{4}\hat{W}^{3}\,+\,\frac{1}{4}\hat{\Phi}_5\,=\,0$.}. The Weyl content of the vector contribution is then given by

\begin{align}\label{vec}
&\int_{0}^{\infty}\frac{dt}{t}\left\{\,\mbox{tr}
\,e^{-(\hat{\Delta}_L^{(1,\bot)}+12)\,t}\,\right\}
\\\nonumber
\\\nonumber
\sim&\int_{0}^{\infty}\frac{dt}{t^{9/2}}\;\frac{e^{-16t}}{7!}
\left\{-252\,\hat{W}^2\cdot t^2\,-
\,\left[\frac{472}{3}\,\hat{W}'^{\,3}\,-\,\frac{40}{3}\,\hat{W}^{3}\,+\,30\,\hat{\Phi}_7\right]\cdot t^3\,+\ldots\right\}
\\\nonumber
\\\nonumber
=&\,\frac{\Gamma(-\frac{1}{2})}{7!}\;\left\{\frac{30368}{3}\,\hat{W}'^{\,3}\,-\,\frac{7904}{3}\,\hat{W}^{3}\,
+\,2568\,\hat{\Phi}_7\,+\ldots\right\}
\end{align}\\
The ellipsis stands for an infinite number of pointwise Weyl invariants of higher order, starting with quartic contractions of the bulk Weyl tensor $\hat{W}^{\,4}$, that do not contribute to the holographic Weyl anomaly. This is the essential feature of the alleged WKB exactness: after factorization of $e^{-16 t}$ the $\hat{W}^{\,2}$ only appears accompanying $t^2$, and the three $\hat{W}'^{\,3}, \hat{W}^{\,3}$ and $\hat{\Phi}_7$ only appear accompanying $t^3$; these are the only bulk Weyl invariants that contribute to the holographic Weyl anomaly, i.e., to the IR logarithmic divergence or the pole in dimensional regularization.\\

The case of the transverse-traceless symmetric rank-two tensor requires more effort, but we have not succeeded in finding any explicit computation in generic dimension $d$ nor in 7D. We have to subtract the longitudinal vector and the trace scalar (cf. appendix~\ref{app.B}). To simplify the computation of the $\hat{\bar{b}}_6$ coefficient we perform a constant shift and leave only the Weyl tensor in the endomorphism $-\hat{\nabla}_2^2-2\hat{W}$.

\begin{equation}\label{cons}
\mbox{tr}\,e^{-(\hat{\Delta}_L^{(2,\bot\top)}\, + \,12)\,t}\,=\,e^{-9t}\left\{e^{11t}\,\mbox{tr}\,e^{(\hat{\nabla}_2^2 + 2\hat{W})\,t}
\,-\,e^{-3t}\,\mbox{tr}\,e^{-\hat{\Delta}_L^{(1)}\,t}\,
-\,e^{11t}\,\mbox{tr}\,e^{-\hat{\Delta}_L^{(0)}\,t}\right\}
\end{equation}\\
This time the Weyl content is contained, after convolution with the exponentials, in the following combination of heat coefficients
\begin{align}
\mbox{tr}\,e^{-(\hat{\Delta}_L^{(2,\bot\top)}+12)\,t}\,\sim\;e^{-9t}
&\left\{\hat{\overline{b}}_4\{-\hat{\nabla}_2^2 - 2\hat{W}\}\cdot t^2\,+
\,\left[11\,\hat{\overline{b}}_4\{-\hat{\nabla}_2^2 - 2\hat{W}\}+\hat{\overline{b}}_6\{-\hat{\nabla}_2^2 - 2\hat{W}\}\right]\cdot t^3 \right.
\\\nonumber
\\\nonumber
&-\hat{\overline{b}}_4\{\hat{\Delta}_L^{(1)}\}\cdot t^2\,-
\,\left[-3\,\hat{\overline{b}}_4\{\hat{\Delta}_L^{(1)}\}+\hat{\overline{b}}_6\{\hat{\Delta}_L^{(1)}\}\right]\cdot t^3
\\\nonumber
\\\nonumber
&\left.-\hat{\overline{b}}_4\{\hat{\Delta}_L^{(0)}\}\cdot t^2\,-
\,\left[11\,\hat{\overline{b}}_4\{\hat{\Delta}_L^{(0)}\}+\hat{\overline{b}}_6\{\hat{\Delta}_L^{(0)}\}\right]\cdot t^3\,+\ldots\right\}
\end{align}
\\
We only need to work out the Weyl content of $\hat{\overline{b}}_6\{-\hat{\nabla}_2^2 - 2\hat{W}\}$. The general expression for this heat coefficient can be found in appendix~\ref{app.B}, there are contributions coming from the endomorphism and from the curvature of the connection. When restricted to the PE/E background metric only the traces collected in the table below need to be taken into account\\
\[
\begin{array}{|r c| c| c| c| c|} \hline
& \mbox{Curvature invariant } & \hat{W}'^{\,3} & \hat{W}^{3} &  \hat{\Phi}_7 \\
\hline \mbox{tr}_V\widehat{V}_{1}\quad\vline & \mbox{tr}_V |\hat{\nabla}\hat{F}|^{2} &  -72&18 &-27 \\
\hline \mbox{tr}_V\widehat{V}_{3}\quad\vline & \mbox{tr}_V \hat{F}\hat{\nabla}^2\hat{F} &  72&-18 &27 \\
\hline \mbox{tr}_V\widehat{V}_{4}\quad\vline & \mbox{tr}_V \hat{F}^{3} &  \frac{9}{2}&-\frac{27}{8} &\frac{27}{8} \\
\hline \mbox{tr}_V\widehat{V}_{5}\quad\vline & \mbox{tr}_V \hat{R}iem \hat{F}^{2} &  54&-\frac{45}{2}&\frac{27}{2} \\
\hline \mbox{tr}_V\widehat{V}_{6}\quad\vline & \mbox{tr}_V \hat{R}ic \hat{F}^{2} &  54&-\frac{27}{2}&\frac{27}{2} \\
\hline \mbox{tr}_V\widehat{V}_{7}\quad\vline & \mbox{tr}_V \hat{R} \hat{F}^{2} &  378&-\frac{189}{2}&\frac{189}{2} \\
\hline \mbox{tr}_V\widehat{V}_{9}\quad\vline & \mbox{tr}_V \hat{E}\hat{\nabla}^2\hat{E} &  -24&6&-9 \\
\hline \mbox{tr}_V\widehat{V}_{10}\quad\vline & \mbox{tr}_V |\hat{\nabla}\hat{E}|^{2} &  24&-6&9 \\
\hline \mbox{tr}_V\widehat{V}_{11}\quad\vline & \mbox{tr}_V \hat{E}^{3} &  -8&-1&- \\
\hline \mbox{tr}_V\widehat{V}_{12}\quad\vline & \mbox{tr}_V \hat{E}\hat{F}^{2} &  -12&6 &-3 \\
\hline \mbox{tr}_V\widehat{V}_{16}\quad\vline & \mbox{tr}_V \hat{R}\hat{E}^{2} &  -126&\frac{63}{2} &-\frac{63}{2} \\
\hline
\end{array}
\]
\\
Most of the traces are easily related to the quadratic ones already computed in~\cite{Acevedo:2017vkk}. As for the challenging ones, those of $\hat{V}_4, \hat{V}_{11}$ and $\hat{V}_{12}$, we used the CADABRA code~\cite{Peeters:2006kp,Peeters:2007wn} to implement and compute them. This leads to \\

\begin{equation}
\hat{\overline{{b}}}_4\{-\hat{\nabla}_2^2 - 2\hat{W}\}\,\sim\,~\frac{163}{180}\,\hat{W}^2
\end{equation}\\
\begin{align}
7!\cdot\hat{\overline{{b}}}_6\{-\hat{\nabla}_2^2 - 2\hat{W}\}\,\sim\, \mbox{tr}_V &\bigg\{-3\hat{A}_5 + \frac{14}{3}\hat{A}_{12} - \frac{16}{3}\hat{A}_{15} + \frac{44}{9}\hat{A}_{16} + \frac{80}{9}\hat{A}_{17}
\\\nonumber
\\
\nonumber &+ 14\bigg( 8\hat{V}_1 + 12\hat{V}_3 - 12\hat{V}_4 + 6\hat{V}_5 - 4\hat{V}_6 + 5\hat{V}_7
\\
\nonumber
\\
\nonumber &+ 60\hat{V}_9 + 30\hat{V}_{10} + 60\hat{V}_{11} + 30\hat{V}_{12} + 30\hat{V}_{16}\bigg)\bigg\}
\\\nonumber
\\\nonumber
=&\,-\frac{445256}{9}\,\hat{W}'^{\,3}\,+\,\frac{97244}{9}\,\hat{W}^{3}\,-\,11844\,\hat{\Phi}_7
\end{align}\\
The rest is just a matter of bookkeeping. The Weyl content of the tensor contribution is then given by
\begin{align}\label{tens}
&\int_{0}^{\infty}\frac{dt}{t}\left\{\,\mbox{tr}
\,e^{-(\hat{\Delta}_L^{(2,\bot\top)}+12)\,t}\,\right\}
\\\nonumber
\\\nonumber
\sim&\int_{0}^{\infty}\frac{dt}{t^{9/2}}\;\frac{e^{-9t}}{7!}
\left\{4760\,\hat{W}^2\cdot t^2\,-
\,\left[\frac{6064}{9}\,\hat{W}'^{\,3}\,+\,\frac{12368}{9}\,\hat{W}^{3}\,-\,348\,\hat{\Phi}_7\right]\cdot t^3\,+\ldots\right\}
\\\nonumber
\\\nonumber
=&\,\frac{\Gamma(-\frac{1}{2})}{7!}\;\left\{-\frac{263104}{3}\,\hat{W}'^{\,3}\,+\,\frac{51892}{3}\,\hat{W}^{3}\,
-\,20376\,\hat{\Phi}_7\,+\ldots\right\}
\end{align}\\

We collect now all contributions that remain after the proper-time integrals. To make the matching of divergences more transparent, we include all numerical coefficients in the evaluation of $\log Z^{^{1-loop,+}}_{_{Einstein}}$ and look after the holographic Weyl anomaly (IR logarithmic divergence or pole in dimensional regularization) that we denote by ${\mathcal A}^+$. For the sake of completeness, we include as well the constant term proportional to the volume, that we denote by $\hat{1}$, stemming from pure-Ricci bulk invariants that were already taken into account in the conformally flat case,

\begin{align}
\log Z^{^{1-loop,+}}_{_{Einstein}}\,=\,&\frac{1}{2}\int_{0}^{\infty}\frac{dt}{t}\left\{\mbox{tr}\,e^{-(\hat{\Delta}_L^{(2,\bot\top)}+12)\,t}\,-\,\mbox{tr}
\,e^{-(\hat{\Delta}_L^{(1,\bot)}+12)\,t}\,\right\}
\\\nonumber
\\\nonumber
=\,\frac{-1}{2\,(4\pi)^3\,7!}\int d^6x \sqrt{g_E}&\left\{\,27694080\,\hat{1}\,-\,97824\,\hat{W}'^{\,3}\,+\,19932\,\hat{W}^{3}\,-\,22944\,
\hat{\Phi}_7\,+\ldots\right\}
\end{align}\\
As before, the ellipsis stands for an infinite number of pointwise Weyl invariants of higher order that do not contribute to the holographic Weyl anomaly.\\
The final step to be done is to invoke our simple prescription~\cite{Bugini:2016nvn} and read off the holographic Weyl anomaly ${\mathcal A}^+$. In conformity with our holographic recipe, we simply retain the numerical coefficients in front of each of the 7D bulk curvature invariants $ \hat{W}'^{\,3}, \hat{W}^{3}$ and $\hat{\Phi}_7$ that descend to the corresponding 6D Weyl invariants $I_1, I_2$ and $\Phi_6$, respectively, while  in place of the volume $\hat{1}$ we write down the 6D Q-curvature, more accurately $ -\frac{1}{384}{\mathcal Q}_6$,  which is the celebrated volume anomaly. The overall outcome of this holographic computation is the following
\begin{equation}
  {\mathcal A}^+=-\frac{1}{2}{\mathcal A}
\end{equation}
\\
The direct comparison is possible by making use of the equality
$I_3 \,=\,16\,I_1 - 4\,I_2 + 3\,\Phi_6$, valid on Einstein backgrounds modulo trivial total derivatives. This allows us to display the way in which the usual Weyl anomaly coefficients $a, c_1, c_2$ and $c_3$ are encoded in our alternative basis\\
\begin{align}
{\mathcal A}&=-2{\mathcal A}^+=-48a{\mathcal Q}_6 + (c_1+16c_3+32a)I_1 + (c_2-4c_3-56a)I_2 + (3c_3+24a)\Phi_6
\\\nonumber
\\\nonumber
&=\,-48\,\frac{601}{2016}\,{\mathcal Q}_6 - \frac{2038}{105}\,I_1 + \frac{1661}{420}\,I_2 - \frac{478}{105}\,\Phi_6
\end{align}
\\

In conformity with the AdS/CFT dictionary at one-loop quantum level, we obtain complete agreement between the one-loop boundary UV- and the bulk IR-log divergences of the corresponding dual gravitational theories.\\
Let us be more explicit: according to the holographic formula (eqn.~\ref{holo}), bulk and boundary divergences are related by

\begin{equation}
{\mathcal A}^-\,-\,{\mathcal A}^+\,=\,{\mathcal A}~.
\end{equation}\\
Our result $-2\,{\mathcal A}^+\,=\,{\mathcal A}$
confirms that the bulk divergence with alternate boundary condition $\,{\mathcal A}^-$ is simply equal to $-{\mathcal A}^+$ and the matching is verified. The latter minus sign can be traced back to the dependence on the scaling dimension $(\Delta_+-d/2)$ that changes sign upon continuation to $(\Delta_--d/2)$ (see e.g.~\cite{Giombi:2013yva,Beccaria:2014xda}) and, in the present case, appears squared in the exponent of the proper time integral: a 9 for the rank-two tensor (eqn.~\ref{tens}) and a 16 for the vector ghost (eqn.~\ref{vec}).

\subsection{Miscellany: boundary factorization vs. bulk WKB exactness}
\qquad\\

Before closing, it may be of interest to establish a connection with the holographic expectations based on the functional Schr\"odinger approach~\cite{Mansfield:1999kk,Mansfield:2003bg} (see also~\cite{Liu:2017ruz}), namely, that the holographic Weyl anomaly from the one-loop bulk effective action be
\begin{equation}
{\mathcal A}^+\,=\,-\frac{1}{2}(\Delta_+-3)\,\overline{b}_6~,
\end{equation}\\
in terms of the heat coefficient $\overline{b}_6$ for the corresponding bulk field when restricted to the 6D boundary.
This is indeed a consequence of our more general relation ${\mathcal A}^+=-\frac{1}{2}{\mathcal A}$, restricted to the particular case of a Ricci-flat 6D boundary where the accumulated heat coefficient for the Weyl graviton ($\Delta_+=6$) is simply 3 times that of the Lichnerowicz operator and the accumulated heat coefficient for the ghost vector ($\Delta_+=7$) is 4 times
, as follows from the factorized form of the one-loop partition function~\cite{Beccaria:2017dmw} when restricted to a Ricci-flat background conveniently rewritten in terms of unconstrained vector and traceless rank-two tensor

\begin{equation}
Z^{^{1-loop}}_{_{Weyl}}=\left[\frac{\left(\det{\Delta}_L^{(1)}\right)^{\,4}}
{\left(\det{\Delta}_L^{(2,\top)}\right)^{\,3}}\right]^{1/2}~.
\end{equation}\\
However, we must  reiterate that the validity of this prescription breaks down as soon as the background is no longer Ricci-flat.

Turning to our holographic computation, we first realize that the holographic Weyl anomaly from the bulk $\hat{W}^2$ is precisely the combination ${W}'^{\,3}\,-\,\frac{1}{4}{W}^{3}\,+\,\frac{1}{4}\hat{\Phi}_6$ which happens to vanish on Ricci-flat manifolds. In consequence, the holographic anomaly comes entirely from the $\hat{\overline{b}}_6$ accompanying $t^3$. Upon proper-time integration, under the assumption of WKB exactness, this produces the linear term in $(\Delta_+-3)$ and the divergence matching boils down to the following ``accidental'' equalities between numerical factors of the heat coefficients in 7D (hatted ones) and in 6D (unhatted ones) that can be readily verified~\footnote{This ``accidental'' equalities have previously been noticed for the case of higher spin fields in 5D/4D (cf.~\cite{Acevedo:2017vkk}, footnote 8 therein).}:\\

Spin 2:
\begin{equation}
7!\cdot\hat{\overline{{b}}}_6\{\Delta_L^{(2,\bot\top)}\}=-\frac{9236}{9}\hat{A}_{16}+\frac{18592}{9}\hat{A}_{17}
\end{equation}\\

\begin{equation}
7!\cdot\overline{{b}}_6\{\Delta_L^{(2,\top)}\}=-\frac{9236}{9}{A}_{16}+\frac{18592}{9}{A}_{17}
\end{equation}\\

Spin 1:
\begin{equation}
7!\cdot\hat{\overline{{b}}}_6\{\Delta_L^{(1,\bot)}\}=-\frac{50}{3}\hat{A}_{16}+\frac{112}{3}\hat{A}_{17}
\end{equation}\\

\begin{equation}
7!\cdot\overline{{b}}_6\{\Delta_L^{(1)}\}=-\frac{50}{3}{A}_{16}+\frac{112}{2}{A}_{17}
\end{equation}\\
These equalities, that lend support to the holographic prescription based on the Schr\"odinger approach, are again restricted to boundary Ricci-flatness and provide only partial information on the holographic Weyl anomaly of the one-loop bulk effective action. In any case, an interesting interplay between boundary factorization of the higher derivative kinetic operator and bulk WKB exactness emerges.
\qquad\\
\section{Conclusion}
\qquad\\
Overall, we have unveiled the way in which the UV divergences of the one-loop effective action for the 6D Weyl graviton are encoded in the large-volume asymptotics of the one-loop effective action for the 7D Einstein graviton in an asymptotically AdS background. This matching certainly constitutes a nontrivial and robust test of the AdS/CFT duality at the one-loop quantum level. We recall again the two key features that enabled the unconventional bulk computation: the implied WKB exactness of the heat kernel in the bulk Poincar\'e-Einstein metric, together with the simple recipe of~\cite{Bugini:2016nvn} to read off the holographic anomaly.\\
In addition, it seems worth further studying the curious interplay between the renormalizability, but nonunitarity, of the induced conformal gravity as opposed to the unitarity, but non-renormalizability, of the bulk Einstein gravity.\\
It is also interesting to note that the one-loop computation we have performed here with the aid of background and heat-kernel techniques must certainly have a diagrammatic counterpart. It would be quite desirable to contrast the numerical values of the type-A anomaly coefficient $a$ and the type-B anomaly coefficient $c_3\propto C_T$ with those obtained by the explicit computation of the one-loop Witten diagrams, tadpole and bubble graphs, that produce the CFT one- and two-point correlation function for the stress tensor, a program already initiated by~\cite{Giombi:2017hpr}.\\
To close, let us mention two further directions in which we envisage extensions of our present calculations to higher-derivative operators: one is related to the family of GJMS-like two-tensor operators of~\cite{Mat14}; and the second also starts with the 6D Weyl graviton, but this time it goes higher in spin, i.e., the extension to 6D Conformal Higher Spins~\cite{Tseytlin:2013fca,Acevedo:2017vkk}.
\qquad\\
\qquad\\
\section*{Acknowledgement}
\qquad\\
We wish to thank Y. Matsumoto, K. Mkrtchyan, and A. Tseytlin for valuable discussions. The work of R.A. was supported in part by grant UNAB DI 08-19/REG. D.E.D. acknowledges support by grant UNAB DI 14-18/REG and is also grateful to the Perimeter Institute for Theoretical Physics for the hospitality during the workshop ``Boundaries and Defects in Quantum Field Theory''. Finally, we hope that the present times of unrest we are having in Chile will pass soon and be the prelude for a brighter future.
\qquad\\
\appendix
\section{Second metric variation of the Q-curvature at 6D Einstein background}\label{app.A}
\qquad\\
It is not hard to realize that the action for the 6D conformal gravity under consideration is nothing but the integral of the critical Q-curvature in 6D, the Lagrangian differs only by trivial total derivatives and an unimportant overall numerical factor. The critical 6D Q-curvature is given by~\cite{GovPet02}\\
\begin{align}
Q_6\,=&\,-\,8\,|\nabla P|^2\,-\,16\,P\nabla^2P\,+\,32\,P^3\,-\,16\,PPW
\\
\nonumber
\\
\nonumber &\,+\,16\,JP^2\,-\,8\,J^3\,+\,8\,J\nabla^2J\,-\,\nabla^2\nabla^2J
\end{align}\\
We spare the detailed index contractions for clarity, their explicit form can be found e.g. in~\cite{Bugini:2016nvn}. Our conventions are as follows\\

\begin{equation}
\qquad\qquad\;\;\mbox{Schouten tensor}\qquad (n-2)\,P_{ij}\,=\,R_{ij}-\frac{1}{2(n-1)}R\,g_{ij}
\end{equation}
\begin{equation}
\mbox{Schouten scalar}\qquad J\,=\,P_k^k\,=\,\frac{1}{2(n-1)}R
\end{equation}

\begin{equation}
\mbox{Cotton tensor}\qquad C_{ijk}\,=\,\nabla_k P_{ij} \,-\, \nabla_j P_{ik}
\end{equation}

\begin{align}
\qquad\qquad\quad\mbox{Bach tensor}\qquad B_{ij}\,=&\,\nabla^k C_{ijk} \,+\, P_{kl}W_{i\;j}^{k\;l}\\
\nonumber\\
\nonumber
=&\,\nabla^2 Pij \,-\,\nabla^k\nabla_j P_{ik} \,+\, P_{kl}W_{i\;j}^{k\;l}
\end{align}\\
One can trade away the Weyl tensor contracted with the Schouten tensor to show that the Bach tensor as well as the Q-curvature are both ``pure-Ricci'' by means of the identity
 \begin{equation}
P_{kl}W_{i\;j}^{k\;l}\,=\,-\nabla^k\nabla_iP_{jk}\,+\,\nabla_i\partial_jJ\,+\,n\,P_{ik}P_j^{k}\,-\,P^2g_{ij}
\end{equation}\\

In general, the metric variation of the Q-curvature results in the Fefferman-Graham obstruction tensor $O_{ij}$ in any even dimension~\cite{GraHir05}.
The explicit form of the obstruction tensor can be worked out by hand in lower dimensions, in 4D it is given by the Bach tensor; whereas in 6D it can be conveniently written in terms of the Bach tensor as follows~\cite{GraHir05,GovPet04,FG12} (indices in parenthesis are symmetrized)
\begin{align}
O_{ij}\,=&\,\nabla^2B_{ij} \,+\, 2\,W_{i\;j}^{k\;l}B_{kl} \,-\, 4\,JB_{ij}\,+ \,8P^{kl}\nabla_lC_{(ij)k}\,+ \,4 P_{km}P^m_{\;\;l}W_{i\;j}^{k\;l}
\\
\nonumber
\\
\nonumber &-4\,C^{k\;l}_{\;i}C_{l j k}\,+\,2\,C^{\;k l}_iC_{j k l}\,+\,4\,C_{(ij)}^{\quad l}\nabla_lJ
\end{align}\\
The second metric variation of the Q-curvature, i.e. the first metric variation of the obstruction tensor, when evaluated on a 6D Einstein manifold is greatly simplified by the vanishing of the Bach  $B_{ij}$ and Cotton tensor $C_{ijk}$ and the gradients of the Schouten tensor $P_{ij}=\frac{J}{6}g_{ij}$ and scalar $J=\frac{R}{10}$, so that the only contributions are the following

\begin{align}
\delta O_{ij}\,=&\,\nabla^2\delta B_{ij} \,+\, 2\,W_{i\;j}^{k\;l}\delta B_{kl} \,-\, 4\,J\delta B_{ij}\,+ \,8P^{kl}\nabla_l \delta C_{(ij)k}\,+ \,4\delta\{ P_{km}P^m_{\;\;l}W_{i\;j}^{k\;l}\}
\\
\nonumber
\\
\nonumber =& \,-\Delta_L\delta B_{ij}\,+\, \frac{4}{3}\,J\delta\{B_{ij}\,- \,P_{kl}W_{i\;j}^{k\;l}\}\,+ \,4\delta\{ P_{km}P^m_{\;\;l}W_{i\;j}^{k\;l}\}
\\
\nonumber
\\
\nonumber =& \,-\{\Delta_L\,-\, \frac{4}{3}\,J\}\,\delta B_{ij}
\end{align}\\
Let us now show that the metric variation of the Bach tensor produces a quartic differential operator that when evaluated on an Einstein manifold factorizes into product of two shifted Lichnerowicz operators in any dimension. For this purpose, it is convenient to rewrite the Bach tensor as follows
\begin{equation}
B_{ij}\,=\,\nabla^2 P_{ij} \,-\nabla_j\partial_i J\,+\,2\,W_{i\;j}^{k\;l}P_{kl} \,+\,P^2 g_{ij}\,-\,n\,P_i^k P_{jk}
\end{equation}\\
At an Einstein metric we have $P_{ij}=\frac{1}{n}J g_{ij}$ and vanishing $\nabla J, \nabla P, C$ and $B$. It is then enough to consider only the following variations under the transverse-traceless fluctuation of the gravitational field
\begin{align}
\delta J\,=&\,0\quad , \quad \delta P_{ij}\,=\,\frac{1}{2(n-2)}\{\Delta_L-2J\}\delta g_{ij}\\
\nonumber\\
\delta g^{ij}\,=&\,-\,g^{ik}g^{jl}\delta g_{kl}\quad , \quad g_{kl}\delta W_{i\;j}^{k\;l}\,=\,-W_{i\;j}^{k\;l}\delta g_{kl}\\
\nonumber\\
\delta \Gamma^{\,k}_{i\;j}\,=&\,\frac{1}{2}g^{kl}\{\nabla_i\delta g_{lj}\,+\,\nabla_j\delta g_{il}\,-\,\nabla_l\delta g_{ij}\}
\end{align}\\
We obtain the only non vanishing contributions upon restriction to an Einstein manifold. Omitting obvious indices, the terms into square brackets correspond to the variation of the first, third and last two in the previous expression for the Bach tensor
\begin{align}
\delta B\,=&\,\left[\nabla^2 \delta P \,-\,\frac{J}{n}\nabla^2\delta g\right]\,+\,\left[2\,W \delta P\,-\,2\,W\frac{J}{n}\delta g \right]\,+\,\left[\frac{J^2}{n}\delta g\,-\,2J\delta P\,+\,\,\frac{J^2}{n}\delta g\right]\\
\nonumber
\\
\nonumber
=&\,\{\nabla^2\,+\,2\,W\,-\,2J\}\{\delta P \,-\,\frac{J}{n}\delta g\}\\
\nonumber
\\
\nonumber
=&\,\{-\Delta_L\,+\,2J\}\{\delta P \,-\,\frac{J}{n}\delta g\}
\end{align}\\
We end up then with a general property of the Bach tensor, namely, the factorization of its metric variation on a generic Einstein background into product of two shifted Lichnerowicz Laplacians
\begin{equation}
\delta B_{ij}\,=\,-\frac{1}{2(n-2)}\,\{\Delta_L\,-\, 2\,J\}\,\{\Delta_L\,-\,4\,\frac{n-1}{n}\,J\}\,\delta g_{ij}
\end{equation}\\
Returning to the 6D Q-curvature and obstruction tensor, we finally obtain the factorized form for the transverse-traceless metric fluctuations~\footnote{The longitudinal and trace parts of the metric fluctuations are absent (gauged away) by virtue of the diffeomorphism and Weyl symmetries of the integrated Q-curvature.} of the 6D Q-curvature about a 6D Einstein background\\
\begin{equation}
\delta O_{ij}\,=\,\frac{1}{8}\,\{\Delta_L\,-\, \frac{2}{15}\,R\}\,\{\Delta_L\,-\, \frac{1}{5}\,R\}\,\{\Delta_L\,-\, \frac{1}{3}\,R\}\,\delta g_{ij}
\end{equation}

\qquad\\
\section{Heat kernel coefficient $b_6$ for $-\nabla^2-E$}\label{app.B}
\qquad\\
In this appendix we first write down the general form of the $b_6$ heat coefficient for $-\nabla^2-E$ when acting  on unconstrained tensor fields for general reference and then obtain the corresponding heat coefficient for the transverse-traceless components.\\
We have, according to~\cite{Gilkey:1975iq,Bastianelli:2000hi},

\begin{align}
{b}_6\{-{\nabla}^{2}-E\}\,=&\, \frac{1}{(4\pi)^3 7!}\,\mbox{tr}_V \bigg\{ 18A_1 + 17A_2 -2 A_3 -4A_4 +9A_5
\\
\nonumber
\\
\nonumber & + 28A_6 - 8A_7 + 24A_8 + 12A_9 + \frac{35}{9}A_{10} - \frac{14}{3}A_{11} + \frac{14}{3}A_{12} \\
\nonumber
\\
\nonumber &\, - \frac{208}{9}A_{13} + \frac{64}{3}A_{14} - \frac{16}{3}A_{15} + \frac{44}{9}A_{16} + \frac{80}{9}A_{17}
\\
\nonumber
\\
\nonumber &+ 14\bigg( 8V_1 + 2V_2 + 12V_3 - 12V_4 + 6V_5 - 4V_6 + 5V_7 + 6V_8
\\
\nonumber
\\
\nonumber &+ 60V_9 + 30V_{10} + 60V_{11} + 30V_{12} + 10V_{13} + 4V_{14} \\
\nonumber
\\\nonumber&+ 12V_{15} + 30V_{16} + 12V_{17} + 5V_{18} - 2V_{19} + 2V_{20}\bigg)\bigg\}
\end{align}\\
When restricted to an Einstein metric, one is free to factorize exponentials of the curvature in the heat kernel. That is exactly what we did in order to have the endomorphism given exclusively by the Weyl tensor and to simplify the computation of the traces involved. For the rank-two symmetric tensor, the endomorphism is then $2\,W_{i\;j}^{k\;l}$ and the connection curvature, $(F_{ij})_{kl}^{rs}=2R_{ij\,(k}^{\,\;\;\;\;\;(r}\delta_{l)}^{s)}$.
The relevant traces that remain were listed in section 3.2.\\

In order to consider transverse-traceless metric fluctuations $h_{ij}^{^{\bot \top}}$ we perform the York decomposition (see e.g.~\cite{Percacci} for details)
\begin{equation}
h_{ij}\,=\,h_{ij}^{^{\bot \top}}\,+\,\nabla_i\,\xi_j\,+\,\nabla_j\,\xi_i\,+\nabla_i\nabla_j\sigma\,-\,\frac{1}{n}g_{ij}\nabla^2\sigma
\,+\,\frac{1}{n}g_{ij}h
\end{equation}\\
Let us first show how to work out the heat kernel from that of the unconstrained field for the (Bochner) Laplacian. It is enough to consider the following commutations

\begin{align}
\nabla_2^2\{\nabla_i\,\xi_j\,+\,\nabla_j\,\xi_i\}\,=&\,\nabla_i\{\nabla_1^2\,+\,\frac{n+1}{n(n-1)}R\}\xi_j\,
+\,\nabla_j\{\nabla_1^2\,+\,\frac{n+1}{n(n-1)}R\}\xi_i
\\
\nonumber
\\
\nonumber
\nabla_2^2\{\nabla_i\,\nabla_j\,-\,\frac{1}{n}g_{ij}\nabla_0^2\}\sigma\,=&\,\{\nabla_i\,\nabla_j\,-
\,\frac{1}{n}g_{ij}\nabla_0^2\}\{\nabla_0^2\,+\,\frac{n+1}{n(n-1)}R\}\sigma
\\
\nonumber
\\
\nonumber
\nabla_2^2\{\frac{1}{n}g_{ij}\}h\,=&\,\{\frac{1}{n}g_{ij}\}\{\nabla_0^2\}h
\end{align}\\
Now, since the Weyl tensor is traceless and in an Einstein background it is also divergence free, it commutes with the longitudinal and trace components of the metric fluctuation and we finally get

\begin{equation}
\mbox{tr}\,e^{(\hat{\nabla}_2^2 + 2\hat{W})\,t}\,=\,\mbox{tr}_{_{\bot\top}}\,e^{(\hat{\nabla}_2^2 + 2\hat{W})\,t}
\,+\,\mbox{tr}_{_{\bot}}\,e^{(\hat{\nabla}_1^2 + \frac{n+1}{n(n-1)}R)\,t}\,+\,\mbox{tr}\,e^{(\hat{\nabla}_0^2 + \frac{2}{n-1}R)\,t}+\,\mbox{tr}\,e^{\hat{\nabla}_0^2\,t}
\end{equation}\\
This is slightly different from Lichnerowicz Laplacians that on Einstein spaces intertwine covariant derivatives. To finally connect with eqn.(\ref{cons}) we only need to  notice that the Lichnerowicz Laplacians on an Einstein space acting on scalar, vector and traceless rank-two symmetric tensor take the following simple form\\
\begin{align}
\Delta_L^{(0)}\,=&\,-\nabla_0^2
\\
\nonumber
\\
\Delta_L^{(1)}\,=&\,-\nabla_1^2\,+\,\frac{1}{n}R
\\
\nonumber
\\
\Delta_L^{(2,\top)}\,=&\,-\nabla_2^2\,-\,2\,W\,+\,\frac{2}{n-1}R
\end{align}\\

\qquad\\

\providecommand{\href}[2]{#2}\begingroup\raggedright\endgroup


\begin{thebibliography}{10}

\bibitem{tHooft:1974toh}
G.~'t Hooft and M.~J.~G.~Veltman,
``One loop divergencies in the theory of gravitation,''
Ann.\ Inst.\ H.\ Poincare Phys.\ Theor.\ A {\bf 20} (1974) 69.

\bibitem{Deser:1974cz}
S.~Deser and P.~van Nieuwenhuizen,
``One Loop Divergences of Quantized Einstein-Maxwell Fields,''
Phys.\ Rev.\ D {\bf 10} (1974) 401.

\bibitem{Goroff:1985sz}
M.~H.~Goroff and A.~Sagnotti,
``Quantum Gravity At Two Loops,''
Phys.\ Lett.\  {\bf 160B} (1985) 81.

\bibitem{Stelle:1976gc}
K.~S.~Stelle,
``Renormalization of Higher Derivative Quantum Gravity,''
Phys.\ Rev.\ D {\bf 16} (1977) 953.

\bibitem{Alvarez-Gaume:2015rwa}
L.~Alvarez-Gaume, A.~Kehagias, C.~Kounnas, D.~L\"ust and A.~Riotto,
``Aspects of Quadratic Gravity,''
Fortsch.\ Phys.\  {\bf 64} (2016) no.2-3,  176
[arXiv:1505.07657 [hep-th]].

\bibitem{PvN77}
P.~Van Nieuwenhuizen,
``On the renormalization of quantum gravitation without matter," Annals of Physics 104, no. 1 (1977): 197-217.

\bibitem{Deser:2016tgn}
S.~Deser,
``One-loop gravity divergences in $D>4$ cannot all be removed,''
Gen.\ Rel.\ Grav.\  {\bf 48} (2016) no.12,  157
[arXiv:1609.04432 [gr-qc]].

\bibitem{Pang:2012rd}
Y.~Pang,
``One-Loop Divergences in 6D Conformal Gravity,''
Phys.\ Rev.\ D {\bf 86} (2012) 084039
[arXiv:1208.0877 [hep-th]].

\bibitem{Beccaria:2017lcz}
M.~Beccaria and A.~A.~Tseytlin,
``C$_{T}$ for conformal higher spin fields from partition function on conically deformed sphere,''
JHEP {\bf 1709} (2017) 123
[arXiv:1707.02456 [hep-th]].

\bibitem{GZ03}
C.~R.~Graham and M.~Zworski, ``Scattering matrix in conformal
geometry,'' Invent.\ Math. {\bf 152} (2003) 89
[arXiv:math-DG/0109089].

\bibitem{Bastianelli:2000hi}
F.~Bastianelli, S.~Frolov and A.~A.~Tseytlin,
``Conformal anomaly of (2,0) tensor multiplet in six-dimensions and AdS / CFT correspondence,''
JHEP {\bf 0002} (2000) 013
[hep-th/0001041].

\bibitem{Mat14}
Y.~Matsumoto,
``A GJMS construction for 2-tensors and the second variation of the total Q-curvature,''
Pac. J. Math. {\bf 262} (2013) 437-455
[arXiv:1202.3227[math.DG]].

\bibitem{Liu:2017ruz}
J.~T.~Liu and B.~McPeak,
``One-Loop Holographic Weyl Anomaly in Six Dimensions,''
JHEP {\bf 1801} (2018) 149
[arXiv:1709.02819 [hep-th]].

\bibitem{Giombi:2013yva}
S.~Giombi, I.~R.~Klebanov, S.~S.~Pufu, B.~R.~Safdi and G.~Tarnopolsky,
``AdS Description of Induced Higher-Spin Gauge Theory,''
JHEP {\bf 1310} (2013) 016
[arXiv:1306.5242 [hep-th]].

\bibitem{Gubser:2002zh}
S.~S.~Gubser and I.~Mitra,
``Double trace operators and one loop vacuum energy in AdS / CFT,''
Phys.\ Rev.\ D {\bf 67} (2003) 064018
[hep-th/0210093].

\bibitem{Gubser:2002vv}
S.~S.~Gubser and I.~R.~Klebanov,
``A Universal result on central charges in the presence of double trace deformations,''
Nucl.\ Phys.\ B {\bf 656} (2003) 23
[hep-th/0212138].

\bibitem{Hartman:2006dy}
T.~Hartman and L.~Rastelli,
``Double-trace deformations, mixed boundary conditions and functional determinants in AdS/CFT,''
JHEP {\bf 0801} (2008) 019
[hep-th/0602106].

\bibitem{Diaz:2007an}
D.~E.~Diaz and H.~Dorn,
``Partition functions and double-trace deformations in AdS/CFT,''
JHEP {\bf 0705} (2007) 046
[hep-th/0702163 [HEP-TH]].

\bibitem{Diaz:2008hy}
D.~E.~Diaz,
``Polyakov formulas for GJMS operators from AdS/CFT,''
JHEP {\bf 0807} (2008) 103
[arXiv:0803.0571 [hep-th]].

\bibitem{Diaz:2008iv}
D.~E.~Diaz,
``Holographic formula for the determinant of the scattering operator in thermal AdS,''
J.\ Phys.\ A {\bf 42} (2009) 365401
[arXiv:0812.2158 [hep-th]].

\bibitem{Aros:2009pg}
R.~Aros and D.~E.~Diaz,
``Functional determinants, generalized BTZ geometries and Selberg zeta function,''
J.\ Phys.\ A {\bf 43} (2010) 205402
[arXiv:0910.0029 [gr-qc]].

\bibitem{Dowker:2010qy}
J.~S.~Dowker,
``Determinants and conformal anomalies of GJMS operators on spheres,''
J.\ Phys.\ A {\bf 44} (2011) 115402
[arXiv:1010.0566 [hep-th]].

\bibitem{Aros:2011iz}
R.~Aros and D.~E.~Diaz,
``Determinant and Weyl anomaly of Dirac operator: a holographic derivation,''
J.\ Phys.\ A {\bf 45} (2012) 125401
[arXiv:1111.1463 [math-ph]].

\bibitem{Dowker:2013mba}
J.~S.~Dowker,
``Spherical Dirac GJMS operator determinants,''
J.\ Phys.\ A {\bf 48} (2015) no.2,  025401
[arXiv:1310.5563 [hep-th]].

\bibitem{Giombi:2013fka}
S.~Giombi and I.~R.~Klebanov,
``One Loop Tests of Higher Spin AdS/CFT,''
JHEP {\bf 1312} (2013) 068
[arXiv:1308.2337 [hep-th]].

\bibitem{Tseytlin:2013jya}
A.~A.~Tseytlin,
``On partition function and Weyl anomaly of conformal higher spin fields,''
Nucl.\ Phys.\ B {\bf 877}, 598 (2013)
[arXiv:1309.0785 [hep-th]].

\bibitem{Tseytlin:2013fca}
A.~A.~Tseytlin,
``Weyl anomaly of conformal higher spins on six-sphere,''
Nucl.\ Phys.\ B {\bf 877}, 632 (2013)
[arXiv:1310.1795 [hep-th]].

\bibitem{Giombi:2014iua}
S.~Giombi, I.~R.~Klebanov and B.~R.~Safdi,
``Higher Spin AdS$_{d+1}$/CFT$_d$ at One Loop,''
Phys.\ Rev.\ D {\bf 89} (2014) no.8,  084004
[arXiv:1401.0825 [hep-th]].

\bibitem{Beccaria:2014jxa}
M.~Beccaria, X.~Bekaert and A.~A.~Tseytlin,
``Partition function of free conformal higher spin theory,''
JHEP {\bf 1408} (2014) 113
[arXiv:1406.3542 [hep-th]].

\bibitem{Aros:2014xga}
R.~Aros, F.~Bugini and D.~E.~Diaz,
``On Renyi entropy for free conformal fields: holographic and q-analog recipes,''
J.\ Phys.\ A {\bf 48} (2015) 105401
[arXiv:1408.1931 [hep-th]].

\bibitem{Beccaria:2014xda}
M.~Beccaria and A.~A.~Tseytlin,
``Higher spins in AdS$_{5}$ at one loop: vacuum energy, boundary conformal anomalies and AdS/CFT,''
JHEP {\bf 1411} (2014) 114
[arXiv:1410.3273 [hep-th]].

\bibitem{Beccaria:2014qea}
M.~Beccaria, G.~Macorini and A.~A.~Tseytlin,
``Supergravity one-loop corrections on AdS$_7$ and AdS$_3$, higher spins and AdS/CFT,''
Nucl.\ Phys.\ B {\bf 892} (2015) 211
[arXiv:1412.0489 [hep-th]].

\bibitem{Beccaria:2015vaa}
M.~Beccaria and A.~A.~Tseytlin,
``On higher spin partition functions,''
J.\ Phys.\ A {\bf 48} (2015) no.27,  275401
[arXiv:1503.08143 [hep-th]].

\bibitem{Beccaria:2015uta}
M.~Beccaria and A.~A.~Tseytlin,
``Conformal a-anomaly of some non-unitary 6d superconformal theories,''
JHEP {\bf 1509} (2015) 017
[arXiv:1506.08727 [hep-th]].

\bibitem{Acevedo:2017vkk}
  S.~Acevedo, R.~Aros, F.~Bugini and D.~E.~Diaz,
``On the Weyl anomaly of 4D Conformal Higher Spins: a holographic approach,''
JHEP {\bf 1711} (2017) 082
[arXiv:1710.03779 [hep-th]].

\bibitem{Bugini:2018def}
F.~Bugini and D.~E.~Diaz,
``Holographic Weyl anomaly for GJMS operators: one Laplacian to rule them all,''
JHEP {\bf 1902} (2019) 188
[arXiv:1811.10380 [hep-th]].

\bibitem{Beccaria:2016tqy}
M.~Beccaria and A.~A.~Tseytlin,
``Iterating free-field AdS/CFT: higher spin partition function relations,''
J.\ Phys.\ A {\bf 49} (2016) no.29,  295401
[arXiv:1602.00948 [hep-th]].


\bibitem{Fradkin:1981hx}
E.~S.~Fradkin and A.~A.~Tseytlin,
``Renormalizable Asymptotically Free Quantum Theory of Gravity,''
Phys.\ Lett.\  {\bf 104B} (1981) 377.

\bibitem{Fradkin:1981iu}
E.~S.~Fradkin and A.~A.~Tseytlin,
``Renormalizable asymptotically free quantum theory of gravity,''
Nucl.\ Phys.\ B {\bf 201} (1982) 469.

\bibitem{Bugini:2016nvn}
F.~Bugini and D.~E.~Diaz,
``Simple recipe for holographic Weyl anomaly,''
JHEP {\bf 1704} (2017) 122
[arXiv:1612.00351 [hep-th]].

\bibitem{Beccaria:2017dmw}
M.~Beccaria and A.~A.~Tseytlin,
``C$_{T}$ for higher derivative conformal fields and anomalies of (1, 0) superconformal 6d theories,''
JHEP {\bf 1706} (2017) 002
[arXiv:1705.00305 [hep-th]].

\bibitem{Susskind:1998dq}
L.~Susskind and E.~Witten,
``The Holographic bound in anti-de Sitter space,''
hep-th/9805114.

\bibitem{Percacci}
R.~Percacci,
``An Introduction to Covariant Quantum Gravity and Asymptotic Safety,''
ISSN: 2424-8223, Vol: 3, Page: 1-300, Publication Year 2017.

\bibitem{Camporesi:1993mz}
R.~Camporesi and A.~Higuchi,
``Arbitrary spin effective potentials in anti-de Sitter space-time,''
Phys.\ Rev.\ D {\bf 47} (1993) 3339.

\bibitem{Camporesi:1994ga}
R.~Camporesi and A.~Higuchi,
``Spectral functions and zeta functions in hyperbolic spaces,''
J.\ Math.\ Phys.\  {\bf 35} (1994) 4217.

\bibitem{Gopakumar:2011qs}
R.~Gopakumar, R.~K.~Gupta and S.~Lal,
``The Heat Kernel on $AdS$,''
JHEP {\bf 1111} (2011) 010
[arXiv:1103.3627 [hep-th]].

\bibitem{ISTY99}
C.~Imbimbo, A.~Schwimmer, S.~Theisen and S.~Yankielowicz,
``Diffeomorphisms and holographic anomalies,''
Class.\ Quant.\ Grav.\  {\bf 17} (2000) 1129
[arXiv:hep-th/9910267].

\bibitem{Besse}
A.~Besse,
``Einstein manifolds,''
Springer, 2002.

\bibitem{FG12}
C.~Fefferman and C.~R.~Graham, ``The Ambient Metric.''
Annals of Math. Studies {\bf 178}, Princeton
University Press (2012)
[arXiv:0710.0919[math.DG]].

\bibitem{Peeters:2006kp}
K.~Peeters,
``A Field-theory motivated approach to symbolic computer algebra,''
Comput.\ Phys.\ Commun.\  {\bf 176} (2007) 550
[cs/0608005].

\bibitem{Peeters:2007wn}
K.~Peeters,
``Introducing Cadabra: A Symbolic computer algebra system for field theory problems,''
[hep-th/0701238].

\bibitem{Mansfield:1999kk}
P.~Mansfield and D.~Nolland,
``One loop conformal anomalies from AdS / CFT in the Schrodinger representation,''
JHEP {\bf 9907} (1999) 028
[hep-th/9906054].

\bibitem{Mansfield:2003bg}
P.~Mansfield, D.~Nolland and T.~Ueno,
``Order 1 / N**3 corrections to the conformal anomaly of the (2,0) theory in six-dimensions,''
Phys.\ Lett.\ B {\bf 566} (2003) 157
[hep-th/0305015].


\bibitem{Giombi:2017hpr}
S.~Giombi, C.~Sleight and M.~Taronna,
``Spinning AdS Loop Diagrams: Two Point Functions,''
JHEP {\bf 1806} (2018) 030
[arXiv:1708.08404 [hep-th]].

\bibitem{Tseytlin:2013fca}
A.~A.~Tseytlin,
``Weyl anomaly of conformal higher spins on six-sphere,''
Nucl.\ Phys.\ B {\bf 877} (2013) 632
[arXiv:1310.1795 [hep-th]].

\bibitem{GovPet02}
A.~R.~Gover and L.~J.~Peterson,
``Conformally invariant powers of the Laplacian, Q-curvature, and tractor calculus,''
Commun. Math. Phys. {\bf 235} (2003) 339
[arXiv:math-ph/0201030].

\bibitem{GraHir05}
C.~R.~Graham and K.~Hirachi,
``The ambient obstruction tensor and Q-curvature,'' in {\em AdS/CFT
correspondence: Einstein metrics and their conformal boundaries}, 59-71, IRMA Lect. Math.
Theor. Phys. {\bf 8}, Eur. Math. Soc., Z\"urich, 2005 [arXiv:math/0405068].

\bibitem{GovPet04}
A.~R.~Gover and L.~J.~Peterson,
``The ambient obstruction tensor and the conformal deformation complex,''
Pac. J. Math. {\bf 226} (2006) 309-351
[arXiv:math/0408229].

\bibitem{Gilkey:1975iq}
P.~B.~Gilkey,
``The Spectral geometry of a Riemannian manifold,''
J.\ Diff.\ Geom.\  {\bf 10} (1975) no.4,  601.

\end{thebibliography}
\end{document}